\documentclass{rmf-d}
\usepackage{nopageno,rmfbib,multicol,times,epsf,amsmath,amssymb,cite}
\usepackage[T1]{fontenc} 
\usepackage[]{caption2}
\usepackage{graphicx}
\usepackage{blindtext}
\usepackage{color}
\usepackage{hyperref}
\def\rmfcornisa{Research of Physics \hfill\rmf\ {\bf ??} (*?*) ???--??? \hfill MES? A\~NO?}

\def\rmfcintilla{{\it Rev.\ Mex.\ Fis.\/} {\bf ??} (*?*) (????) ???--???}
\clearpage \rmfcaptionstyle 
\begin{document}
\markboth{Effects of numerical methods on MHD solar wind simulations}{Effects of numerical methods on MHD solar wind simulations}


%
%
\title{Investigating the effects of numerical algorithms on global magnetohydrodynamics (MHD) simulations of solar wind in the inner heliosphere
\vspace{-6pt}}
\author{Luis Ángel de León Alanís}
\address{Facultad de Ciencias Físico Matemáticas, Universidad Autónoma de Nuevo León, Ciudad Universitaria, C.P. 66455 San Nicolás de los Garza, Nuevo León, México \\
e-mail: luis.deleonal@uanl.edu.mx}
%
%
\author{J. J. González-Avilés*}
\address{Escuela Nacional de Estudios Superiores, Unidad Morelia, Universidad Nacional Autónoma de México, C.P. 58190 Morelia, Michoacán, México \\
*Corresponding author: jgonzaleza@enesmorelia.unam.mx}

\author{P. Riley and M. Ben-Nun}
\address{Predictive Science Inc., San Diego, CA 92121, USA \\
e-mail: pete@predsci.com, mbennun@predsci.com}
\maketitle
%
%
\recibido{}{
\vspace{-12pt}}
\begin{abstract}
\vspace{1em} 
%
%

This paper explores the effects of numerical algorithms on global magnetohydrodynamics (MHD) simulations of solar wind (SW) in the inner heliosphere. To do so, we use sunRunner3D, a 3-D MHD model that employs the boundary conditions generated by CORHEL and the PLUTO code to compute the plasma properties of the SW with the ideal MHD approximation up to 1.1 AU in the inner heliosphere. Mainly, we define three different combinations of numerical algorithms based on their diffusion level. This diffusion level is related to the way of solving the MHD equations using the finite volume formulation, and, therefore, we set in terms of the divergence-free condition methods, Riemann solvers, variable reconstruction schemes, limiters, and time-steeping algorithms. According to the simulation results, we demonstrate that sunRunner3D reproduces global features of Corotating Interaction Regions (CIRs) observed by Earth-based spacecraft (OMNI) for a set of Carrington rotations that cover a period that lays in the late declining phase of solar cycle 24, independently of the numerical algorithms. Moreover, statistical analyses between models and in-situ measurements show reasonable agreement with the observations, and remarkably, the high diffusive method matches better with in-situ data than low diffusive methods.

\vspace{1em}
\end{abstract}
\keys{ \bf{\textit{
Solar wind; Magnetohydrodynamics (MHD); numerical methods; heliosphere
}} \vspace{-8pt}}

\section{Introduction}
\label{Intro}

The Sun continuously emits charged particles in the state of plasma that disperse throughout the solar system, forming what we know as the solar wind (SW). This phenomenon originates in the solar corona, where not all the mass is in balance, which causes the release of part of it and its subsequent dispersion into space. Since this set of charged particles is in motion, it carries a magnetic field that forms the heliosphere. This magnetic field protects the planets and other celestial objects in the solar system from the influence of the interstellar wind and high-energy particles from outer space.

Knowledge about the plasma's physical properties that constitute the SW is essential to understanding the connection between the Sun and the Earth, i.e., the Space Weather. The physical properties of the SW have been studied using in situ observations, e.g., the Advanced Composition Explorer (ACE), WIND, and the Deep Space Climate Observatory (DSCOVR) covering the near-Earth region. Moreover, the Parker Solar Probe (PSP) and Solar Orbiter (SolO) observe the properties of the inner heliosphere. All the above observatories provide in-situ measurements of physical properties, such as velocity, proton density, temperature, and magnetic field of the SW. However, they do not accurately estimate key forecasting parameters, such as arrival times of SW currents near the Earth, i.e., at about 1 AU. Therefore, many current investigations employ numerical models to improve these limitations to develop a broader picture of space weather events, such as SW streams, Corotating Interaction Regions (CIRs), and Stream Interaction Regions (SIRs). 

Specifically, there is a more noticeable advance in numerical models applied to study the propagation and dynamics of the SW in the inner heliosphere. The numerical models are based on photospheric magnetograms and numerically solving the MHD equations to describe the propagation and evolution of the SW streams in the inner heliosphere globally. Among these models are, for example, MAS \cite{Riley_et_al_2001}, ENLIL \cite{2003AdSpR..32..497O}, SWWF \cite{Toth_et_al_2005}, SIP-CESE \cite{Feng_et_al_2010}, SUSANOO \cite{Shiota_et_al_2014}, EUHFORIA \cite{Pomoell&Poedts_2018}, and more recently, SWASTi-SW \cite{Mayank_2022}. Furthermore, MHD simulations have been used to model the solar wind to analyze its behavior, including its turbulence ~\cite{1}, their interaction and their effects on the Earth's magnetosphere ~\cite{2,3}, as well as in the magnetosphere of other planets, such as Mercury ~\cite{4} or Jupiter ~\cite{5}. However, despite all these models, there are not many investigations, aside from \cite{2023SunGe..15...49B,2016FrASS...3....6Z}, that explore the potential effects of numerical methods associated with the solution of MHD equations in 3D spherical coordinates applied to simulate steady-state solutions of the SW streams in the inner heliosphere.

In this paper, we use sunRunner3D \cite{riley23b} to investigate the effects of the numerical algorithms on global 3D MHD simulations of steady-state solutions of SW in the inner heliosphere. To achieve this, we define three different combinations of numerical algorithms based on their diffusion level. The numerical algorithms are related to solving the MHD equations using the finite volume formulation and, therefore, can be defined in terms of the divergence-free condition methods, Riemann solvers, variable reconstruction schemes, limiters, and time-steeping algorithms. We organize the paper as follows: in section \ref{Model}, we describe the sunRunner3D model and the parameters space; in section \ref{Results_of_num_simulations}, we show the results of the numerical simulations for a set of Carrington Rotations, the comparisons of the model results with OMNI 1-hr in-situ measurements and the statistical analysis. Finally, in section \ref{conclusions}, we draw our conclusions and final comments.   

\section{Model}
\label{Model}

To perform the simulations of the steady-state SW solutions of this paper, we employ sunRunner3D, which is a community-developed open-source package and an improvement of sunRunner1D \cite{Riley&Ben-Nun_2022}, which is a tool for exploring ICME evolution through the inner heliosphere, considering spherical symmetry in 1D. SunRunner3D has been applied to interpret the global structure of the heliosphere from in situ measurements \cite{Gonzalez-Aviles_et_al_2023} and to globally simulate a set of SIRs/CIRs observed by the Parker Solar Probe (PSP) Mission during its first five orbits and by the STEREO-A (STA) mission \cite{Aguilar-Rodriguez_et_al_2023}. In particular, sunRunner3D, in the coronal domain, uses the boundary conditions generated by CORona-HELiosphere (CORHEL) \cite{2009AGUFMSA43A1612L}. In contrast, in the inner heliosphere domain, it employs the PLUTO code \cite{Mignone_et_al_2007} to compute the plasma properties of SW with the magnetohydrodynamic (MHD) approximation up to 1.1 AU in the inner heliosphere. 

\subsection{CORHEL}
\label{CORHEL}

To generate the boundary conditions, we used the CORHEL (CORona-HELiosphere; \cite{Linker_2016, Riley_et_al_2012}) framework, which is capable of modeling the ambient solar corona and the inner heliosphere for a specific period of interest. Mainly, it derives the boundary conditions using maps of the Sun's photospheric magnetic field derived from magnetograms. These magnetograms are obtained principally from SDO's HMI instrument. Then, it runs the coronal model using the MAS code \cite{1996AIPC..382..104M,1999PhPl....6.2217M}) until obtaining a relaxed state, which serves to generate the boundary conditions for the heliospheric models. The CORHEL solutions used in this paper are available at the Predictive Science website (\href{https://www.predsci.com/data/runs/}{https://www.predsci.com/data/runs/}). There, the boundary conditions are already in a readable format for the MHD code; then, we use them to drive the inner heliospheric MHD model, described in the following subsection. 

\subsection{MHD model}
\label{MHD_model}

The inner heliosphere model employs the PLUTO code \cite{Mignone_et_al_2007} from $R_{b}=0.14$ AU outwards to solve the three-dimensional time-dependent MHD equations in spherical coordinates. Specifically, we adopt the ideal MHD equations written in the following dimensionless conservative form,

\begin{eqnarray}
\frac{\partial\varrho}{\partial t} + \nabla\cdot(\varrho{\bf v}) &=& 0, \label{density}\\
\frac{\partial(\varrho{\bf v})}{\partial t} + \nabla\cdot(\varrho{\bf v}{\bf v}-{\bf B}{\bf B} + p_{t}{\bf I}) &=& {\bf F},  \label{momentum} \\
\frac{\partial E}{\partial t} +\nabla\cdot((E+p_{t}){\bf v}-{\bf B}({\bf v}\cdot{\bf B})) &=& {\bf v}\cdot{\bf F}, \label{energy} \\
\frac{\partial{\bf B}}{\partial t} +\nabla\cdot({\bf v}{\bf B} -{\bf B}{\bf v}) &=& 0, \label{evolB} \\
\nabla\cdot{\bf B} = 0, \label{divB}
\end{eqnarray}

\noindent where $\varrho$ is the gas density, ${\bf v}$ represents the fluid velocity, $E$ is the total energy density, ${\bf B}$ is the magnetic field, $p_{t}$ is the total pressure (thermal+magnetic), and ${\bf I}$ is the unit matrix. Specifically, $p_{t}=p+{\bf B}^{2}/2$, where by the ideal gas law $p= \varrho k_{B} T/ {\bar{m}}$. Here $T$ is the temperature of the plasma, $\bar{m}=\mu m_{H}$ is the particle mass specified by a mean molecular weight value $\mu = $0.6, which is a typical value for a fully ionized gas primarily consisting of hydrogen with a small component of helium, $m_{H}$ is the mass of the hydrogen atom, and $k_{B}$ is the Boltzmann constant. In addition, $E= p/(\gamma-1) + \varrho{\bf v}^{2}/2 + B^{2}/2$, being $\gamma=5/3$ the polytropic index. The source terms $\bf F$ represent the Coriolis and centrifugal forces conservatively \cite{1998A&A...338L..37K, Mignone_et_al_2012}.

\subsubsection{Parameters}
\label{parameter_space}

To investigate the effects of the numerical algorithms, we chose three models in terms of combinations of Riemann solvers, variable reconstruction schemes, limiters, and time-integration schemes. The three models' specific combinations are classified in terms of the diffusion level as shown in Table \ref{Tab:1}. For example, Model 1 employs the Riemann solver Roe \cite{ROE1981357}, the LimO3 \cite{2009JCoPh.228.4118C}, which is a compact stencil third-order reconstruction scheme, the minmod limiter, and the RK3 time-stepping algorithm. For Model 2, we chose the Harten–Lax–van Leer–Contact (HLLC) Riemann solver \cite{Li_2005}, the WENO3  scheme, which provides third-order weighted essentially non-oscillatory reconstruction \cite{2009JCoPh.228.4248Y}, VanLeer-limiter and the RK2 time-stepping algorithm, which represents a medium level of diffusivity. Finally, for Model 3, we select the Total Variation Diminishing Lax Friedrichs scheme (TVDLF; \cite{YEE1987151}), a second-order linear reconstruction scheme in combination with the minmod limiter and the RK2 time-stepping algorithm for the time integration; this model represents the maximum level of diffusivity. For the three models, to ensure the divergence-free condition (Eq. [\ref{divB}]), we selected the Powell eight-waves formulation \cite{Powell1994}, which we denote as the Powell method from now on, and the hyperbolic divergence cleaning method \cite{DEDNER2002645}, which we call HDC throughout the paper. According to the aforementioned, Model 1 is the low-diffusive method since it uses higher-order algorithms. Model 2 is the medium diffusive combination, employing a lower-order Riemann solver and time-stepping algorithm. Model 3 is highly diffusive since it includes the lowest-order algorithms. For HDC, Model 1 employs the Harten–Lax–van Leer–Discontinuities (HLLD) Riemann solver \cite{2005JCoPh.208..315M} instead of Roe. This choice is based on stability reasons, but HLLD is typically more robust and efficient than the linearized Riemann solver such as Roe, and it is also a low diffusive solver. Finally, we do not use HLLD combined with the Powell method since they are incompatible.  

\begin{table*}
\label{Tab:1}
\centering
\begin{tabular}{|c|c|c|c|c|c|}
\hline
    Model     & Riemann Solver & Reconstructor + Limiter + Time-stepping & Divergence-free method & Diffusion level       \\ \hline
1 & Roe (HLLD)   & LimO3 + MINMOD + RK3   & Powell method (HDC) &    Low        \\ \hline
2 & HLLC   & WENO3 + VANLEER  + RK2 &  Powell method \& HDC &   Medium        \\ \hline
3 & TVDLF  & LINEAR + MINMOD + RK2  &  Powell method \& HDC    &  High        \\ \hline
\end{tabular}
\caption{Numerical algorithms of the simulation models}
\end{table*}

\section{Results of the numerical simulations}
\label{Results_of_num_simulations}

We select the following Carrington rotations (CRs) to investigate the effects of the numerical algorithms: CR2215, CR2220, CR2230, CR2250, and CR2260. These CRs are between March 2019 and August 2022, coinciding with the decreasing phase (solar minima conditions) of cycle 24 and the early ascending phase of cycle 25. Notably, in CR2250, a transient event related to a Coronal Mass Ejection (CME) was detected; therefore, for this CR, the results of the ambient SW solutions could not be accurate.       

\begin{figure*}
 \centering
 \includegraphics[width=0.47\linewidth]{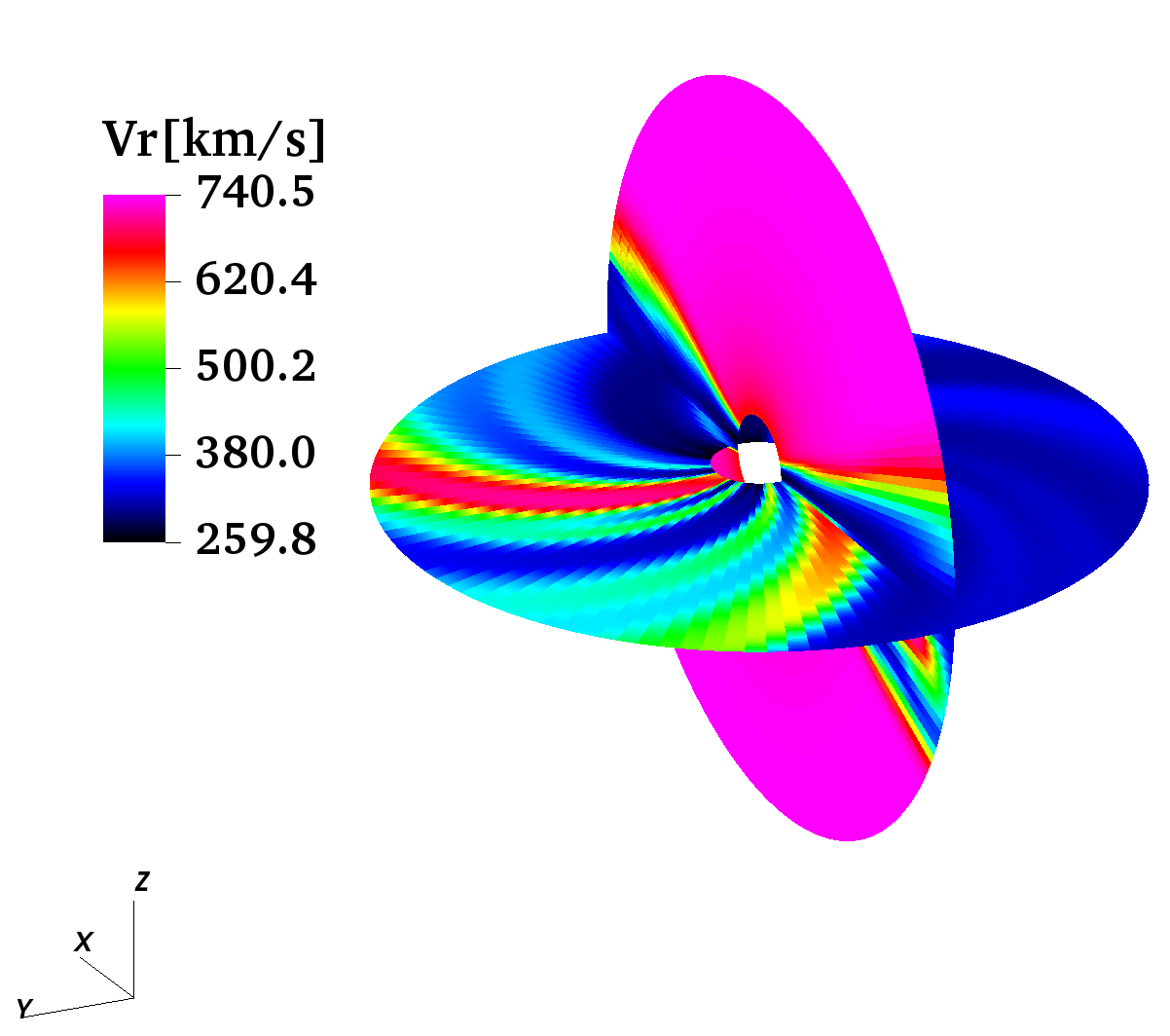}
 \includegraphics[width=0.47\linewidth]{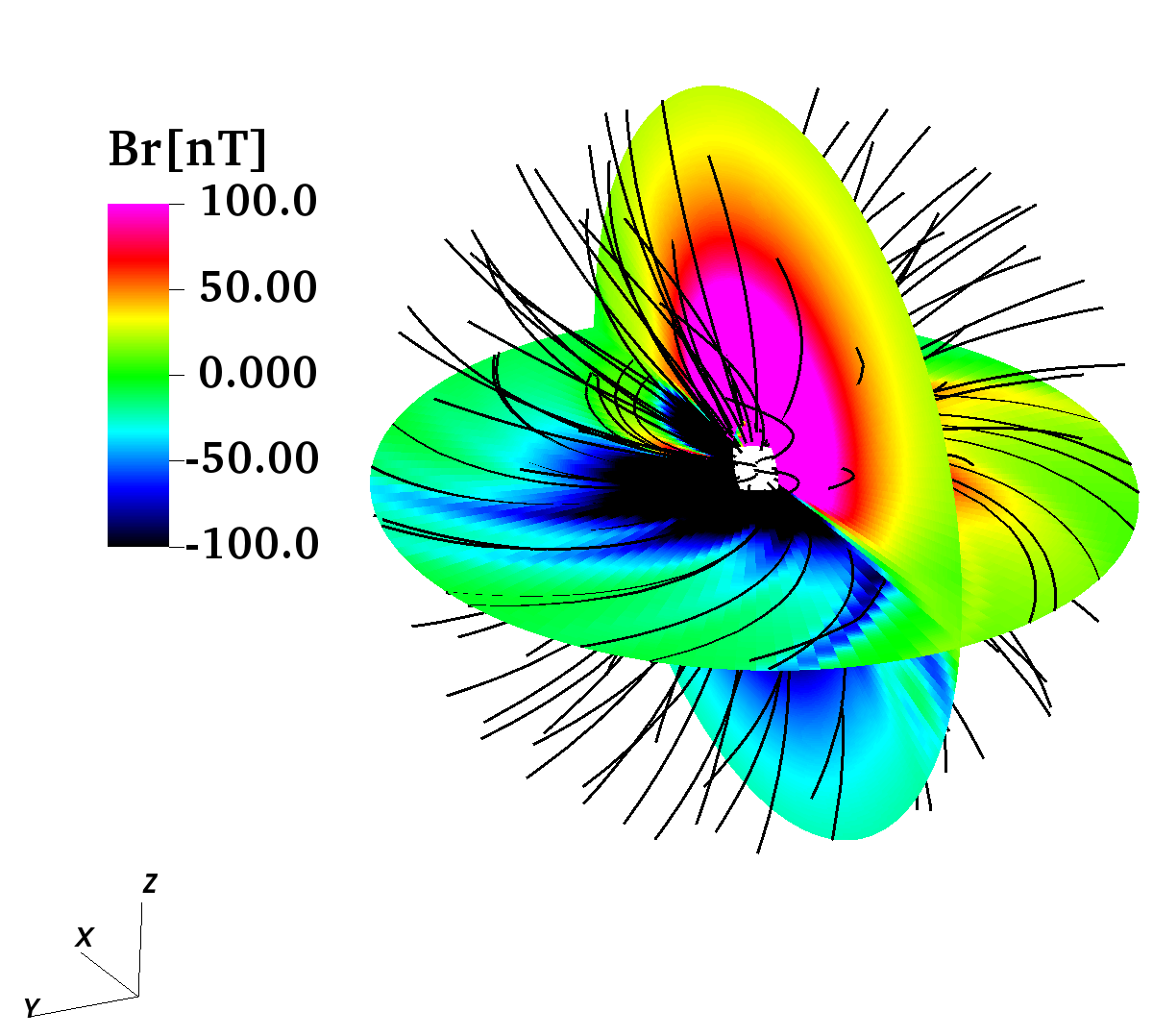}
 \caption{Maps of simulated SW parameters in the heliospheric equatorial and meridional planes for CR2215. We show velocity $V_{r}$ (km s$^{-1}$) at the left, and at the right, we display the radial magnetic field $B_{r}$ (nT), overlay with magnetic field lines in black color.}
    \label{fig:solar_wind_relaxation_3D_CR2215_Model_1}
\end{figure*}

In Figure \ref{fig:solar_wind_relaxation_3D_CR2215_Model_1}, we show the results for the relaxation of the SW conditions corresponding to CR2215 using the solution of Model 1 for the Powell method. For this simulation, we ran sunRunner3D long enough to convect any/all transient phenomena created at $t=0$ past the outer radial boundary. In particular, we display equatorial and meridional cuts, observing standard features of interplanetary solutions expected for the steady state SW. We recognize the fast and slow SW streams in the radial velocity as CIRs in steady-state SW conditions. Notably, in the plot of the radial velocity $V_{r}$, it is also visible that the high-speed ($>700$ km s$^{-1}$) SW dominates in the north and south poles, as shown in the meridional plane. We also observe a mix of slow and fast wind at all latitudes. Finally, in the radial magnetic field cuts, we identify the formation of the Parker spiral represented by the magnetic field lines colored in black. 

\begin{figure*}
 \centering
 \includegraphics[width=0.47\linewidth]{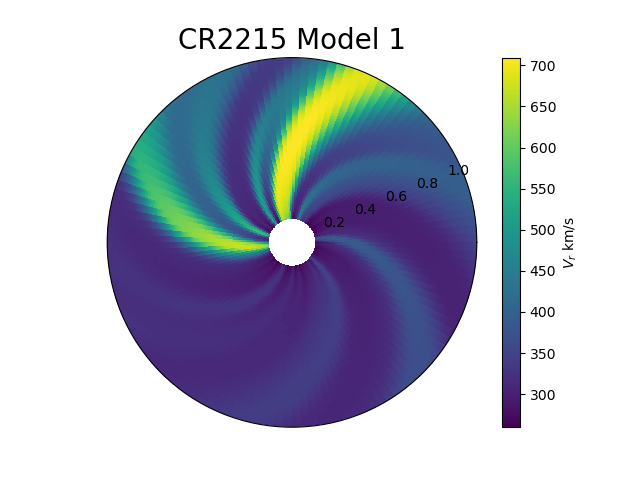}
 \includegraphics[width=0.47\linewidth]{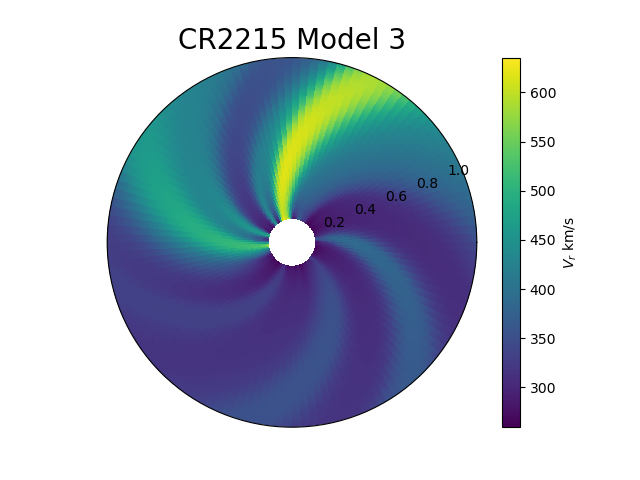}
  \includegraphics[width=0.47\linewidth]{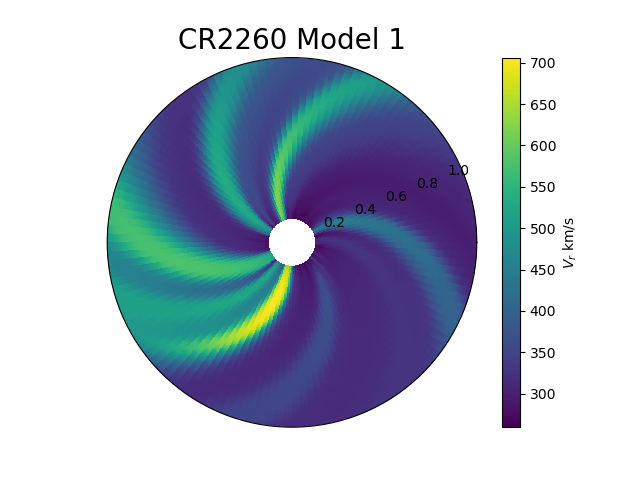}
   \includegraphics[width=0.47\linewidth]{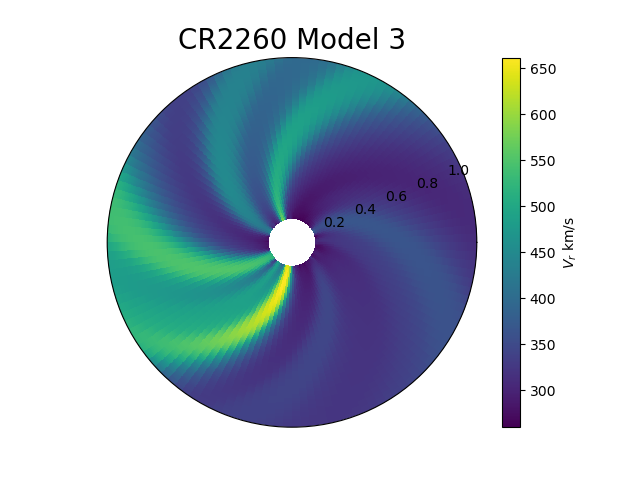}
 \caption{Snapshots of the radial velocity $V_{r}$ (km s$^{-1}$) for the relaxed solar wind solutions in the equatorial plane corresponding to CR2215 (top) and CR2260 (bottom), for Model 1 (left panels) and Model 3 (right panels).}
    \label{fig:solar_wind_relaxation_solutions}
\end{figure*}

Besides, in Figure \ref{fig:solar_wind_relaxation_solutions}, we show the results of the radial velocity $V_{r}$ in km s$^{-1}$ for the relaxation of the SW conditions corresponding to CR2215 and CR2260 for Models 1 and 3 using the Powell method. In the four panels, we display radial velocity cuts in the equatorial plane obtained with the PsiPy \cite{David_Stansby_and_Pete_Riley_PsiPy} tool, which is helpful for reading, processing and visualizing MHD models developed by Predictive Science Inc. Particularly in these cuts, we observe the typical structures, such as slow and fast SW streams, typically denominated as CIRs. We also note that the solutions for Model 1 show the SW streams well structured, and the solutions for Model 3 look more diffusive than Model 1, but they consistently reproduce the SW streams' global structure.

\subsection{Comparisons with in-situ measurements}
\label{comparisons_with_in-situ_measurements}

To identify the differences between the solutions of the models listed in Table \ref{Tab:1} more clearly, we compare the steady-state SW solutions obtained with sunRunner3D with in-situ measurements from OMNI (Earth-based spacecraft) for all the CRs listed above. To do so, we use the PsiPy tool, which can also compare in-situ data with sunRunner3D model results. Notably, for the CRs studied in this paper, the position of both spacecraft is around the L1 Lagrange point, i.e., more than 0.394 solar radii upstream of Earth. Therefore, the spacecraft's location was not close to the magnetopause since its distance from the Earth's center to the "nose" is about 0.096 solar radii and to the flanks abreast of the Earth about 0.137 solar radii. In comparison, the radius of the distant tail is 0.229-0.275 solar radii.

\begin{figure*}
\centering
 \includegraphics[width=0.72\linewidth]{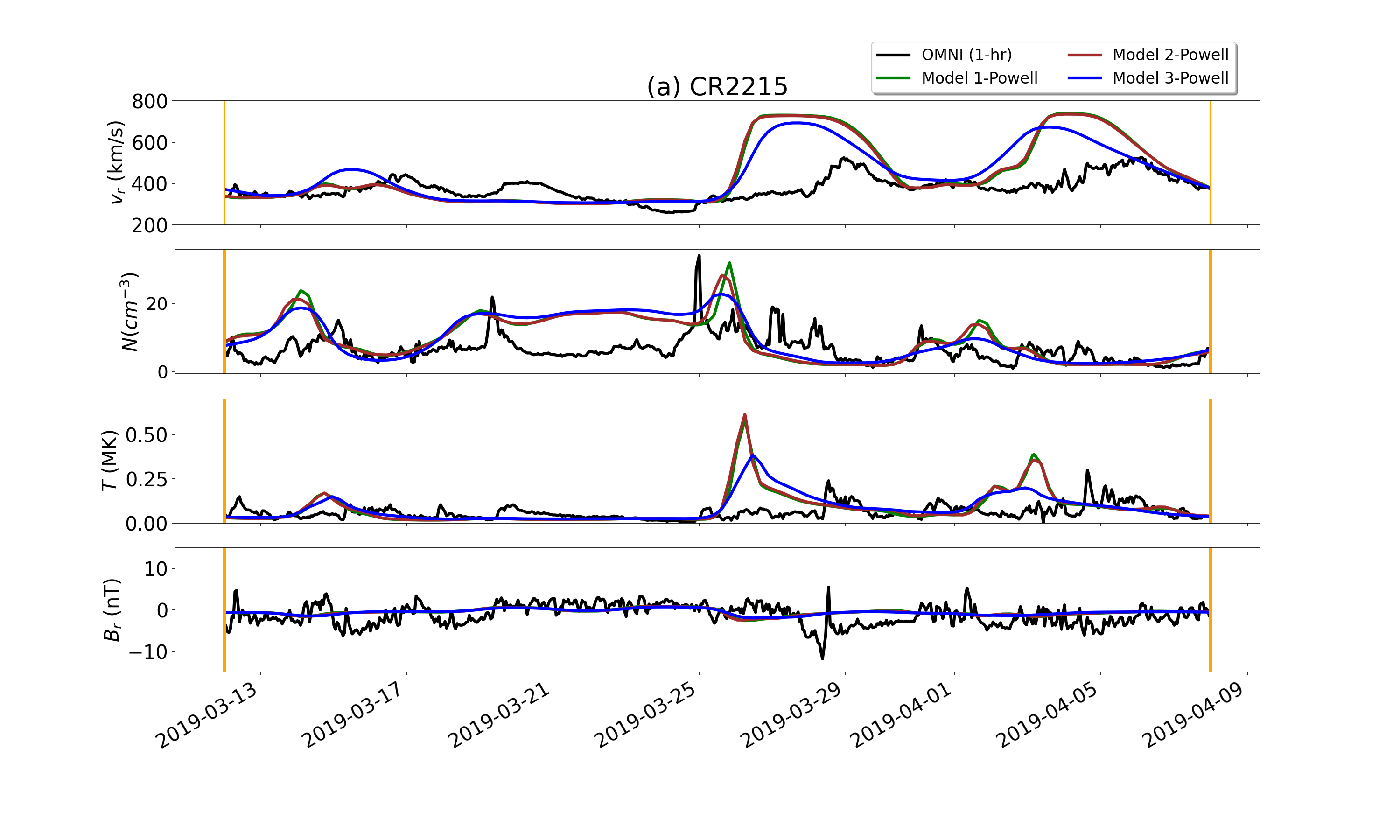}
    \includegraphics[width=0.72\linewidth]{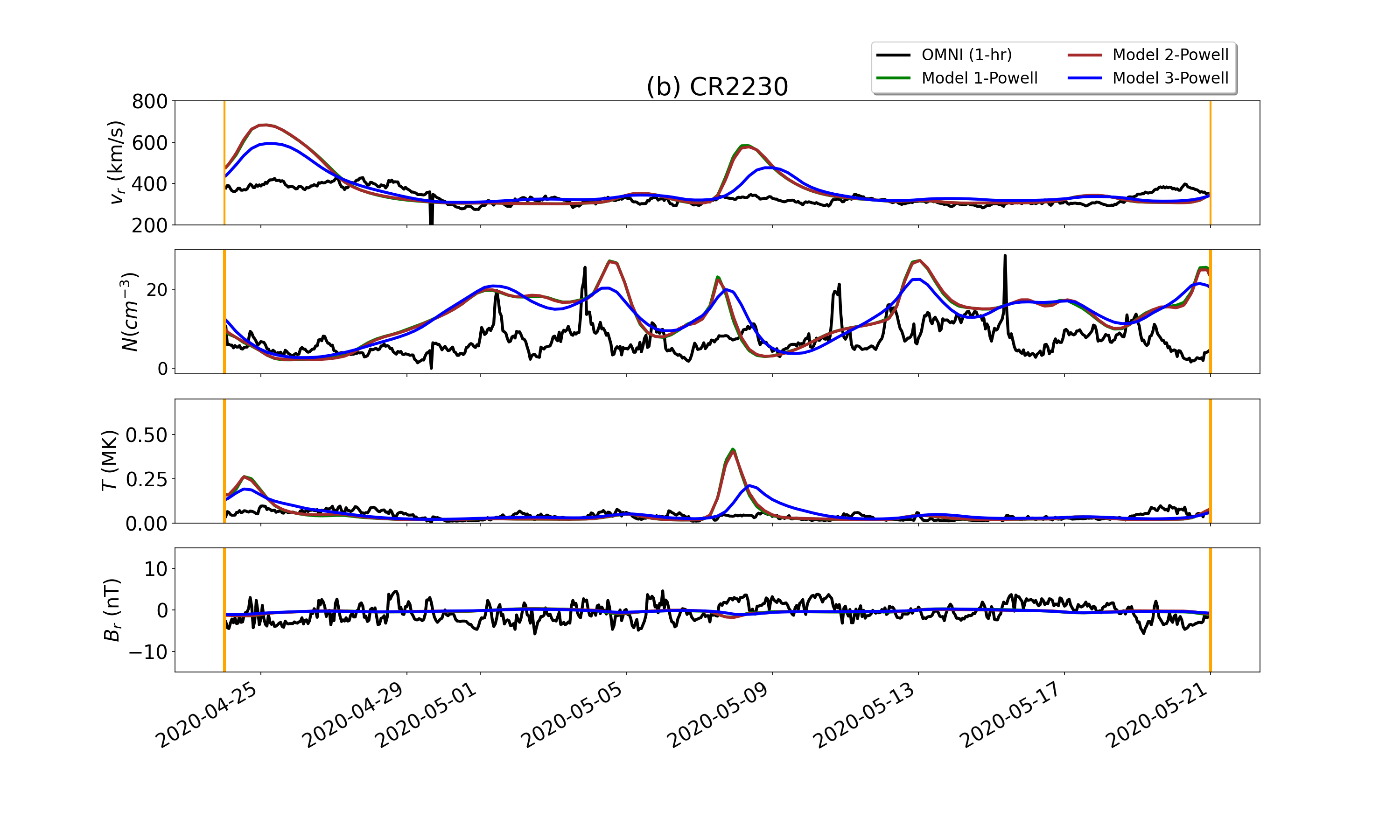}
    \includegraphics[width=0.72\linewidth]{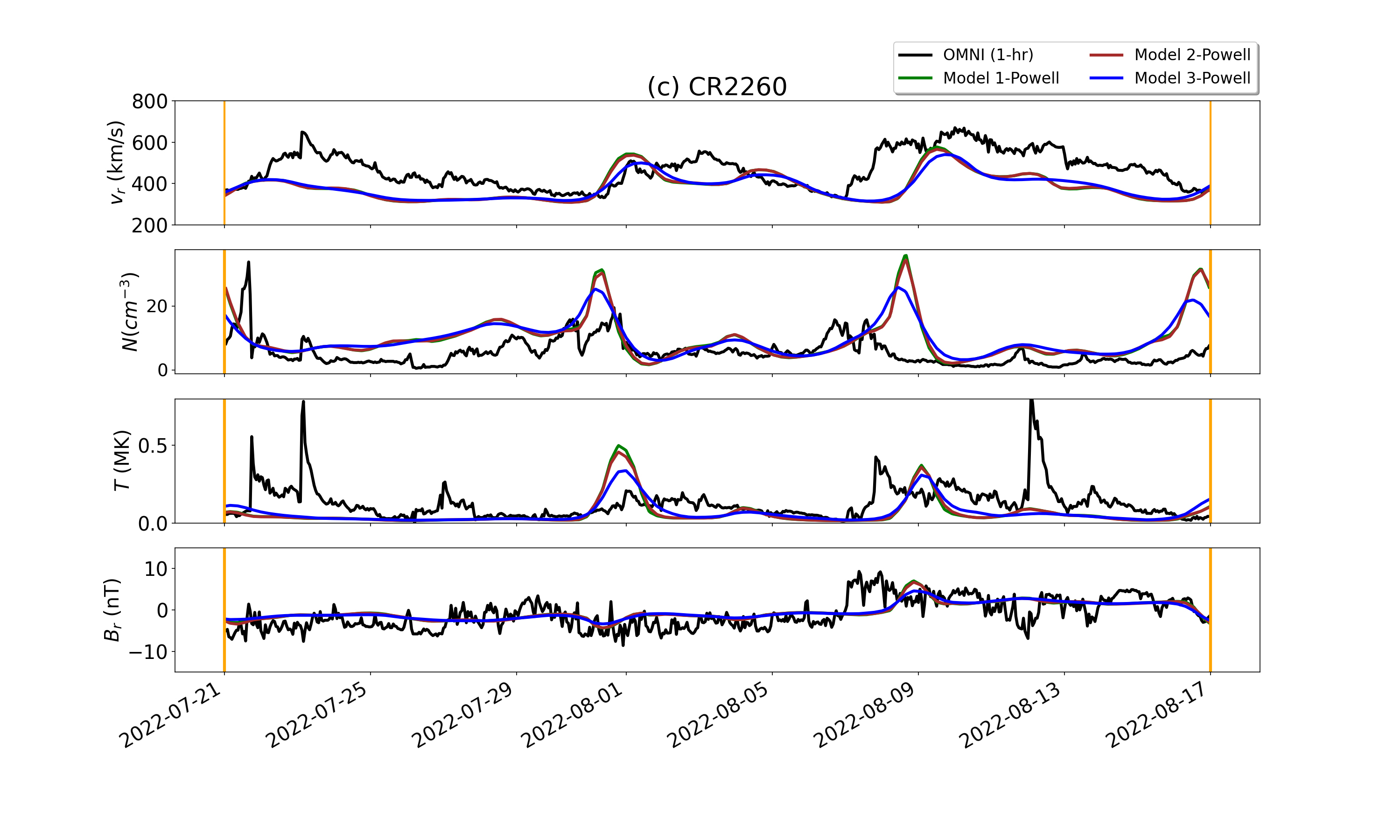}
     \caption{Comparison of Model 1 (green curves), Model 2 (brown curves), and Model 3 (blue curves) solutions for the Powell method corresponding to the radial velocity $V_{r}$ (km s$^{-1}$), radial magnetic field $B_{r}$ (nT), number density $N$ (N/cm$^{3}$), and temperature $T$ (MK) with OMNI 1-hr in-situ measurements (black curves) for CR2215 (a), CR2230 (b), and CR2260 (c).}
    \label{fig:in-situ_comparisons_OMNI}
\end{figure*}

In Figure \ref{fig:in-situ_comparisons_OMNI}, we show the results for comparisons between the steady state SW solutions of Model 1 (green curves), Model 2 (brown curves), and Model 3 (blue curves) using the Powell method for the divergence-free condition with the OMNI 1-hr in-situ measurements (black curves) for three representative CRs: CR2215, CR2230, and CR2260. For example, the comparisons between the models and the radial velocity $V_{r}$ for CR2215 show regions of slow SW ($\sim 300$ km s$^{-1}$) in the first twelve days observed by OMNI, which the three models globally capture. However, the three models overestimate the two rises of the SW speed from 300 km s$^{-1}$ to 500 km s$^{-1}$ observed on 2019-04-01 and 2019-04-08. As expected, Model 3 seems more diffusive than Models 1 and 2. Additionally, we see that Model 2 and Model 3 give similar solutions. In the case of the radial magnetic field, we see that the three models are similar and only capture global variations, i.e., the changes of sign, but underestimate their amplitude.
Regarding number density, the three models overestimated the observed values of OMNI, but they behave similarly despite the algorithmic combination. It is also evident in the number density that Model 3 is more diffusive than Models 1 and 2 since it shows lower densities at the sharply observed regions. For the comparisons with temperature, we see that the three models consistently capture the regions of low temperatures; however, the models overestimate the regions with high temperatures. Also, the most diffusive solution is given by Model 3. The global behavior is consistent with the results shown in the number density. 

Regarding the results of the comparisons between models with OMNI in-situ measurements for CR2230, we note a similar behavior as the results for CR2215. Remarkably, three models overestimated temperature, and the most diffusive solution was achieved by Model 3. In the case of CR2260, we show that three models underestimated the rise in speed. However, they globally matched the number density and, again, underestimated the magnetic field strength of the radial field. The highly diffusive solution is shown for Model 3, but Models 1 and 2 are similar in all the variables for these three representative CRs. 


\begin{figure*}
\centering
 \includegraphics[width=0.72\linewidth]{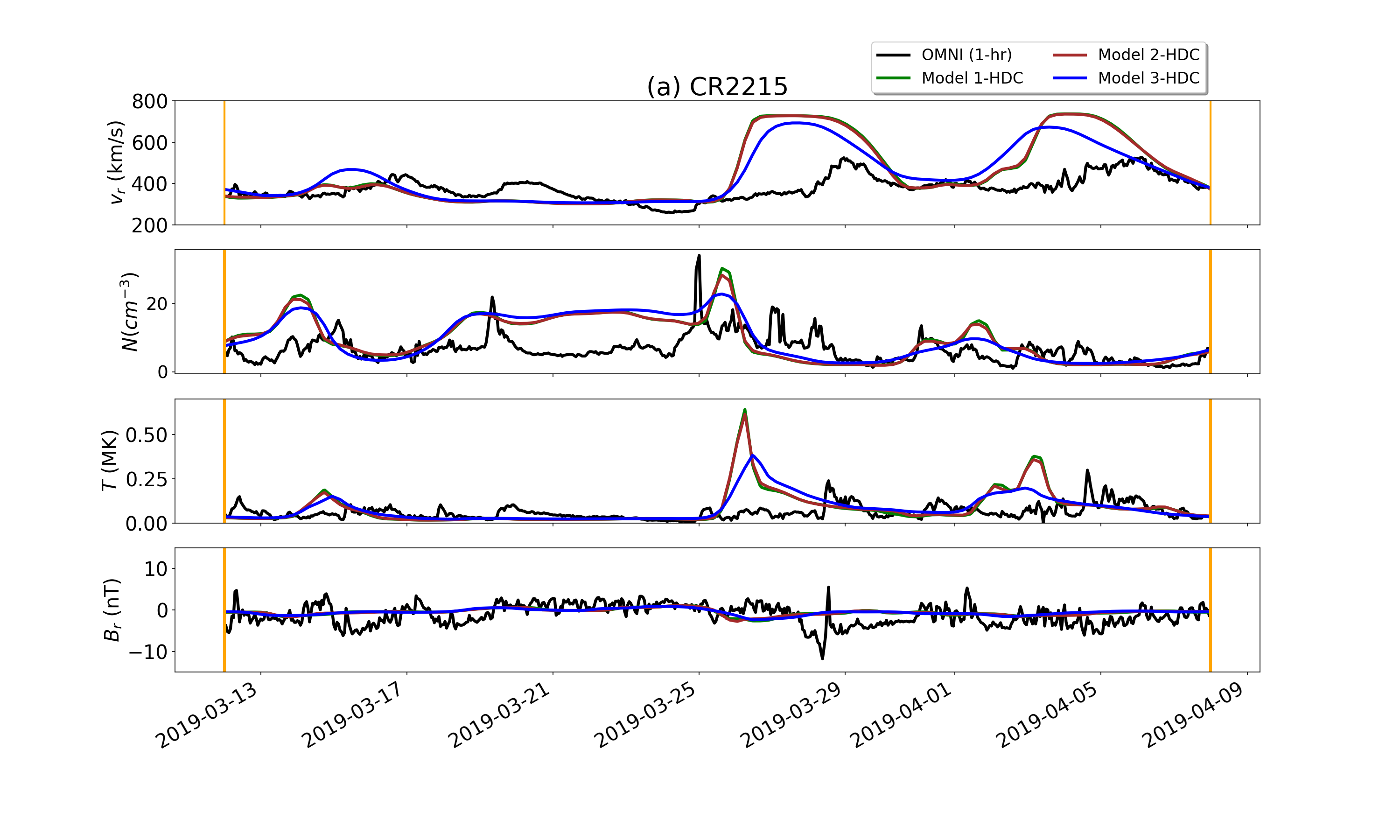}
    \includegraphics[width=0.72\linewidth]{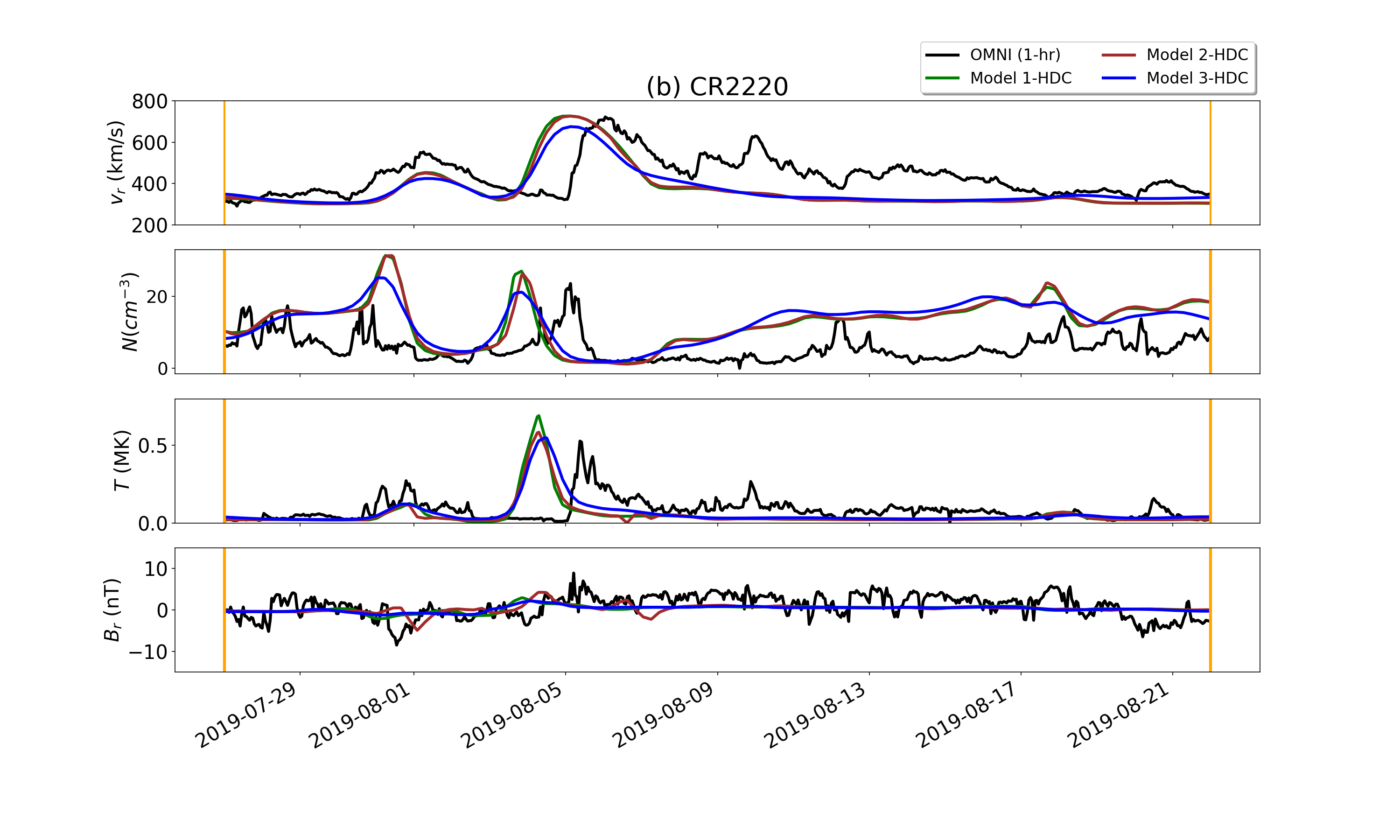}
    \includegraphics[width=0.72\linewidth]{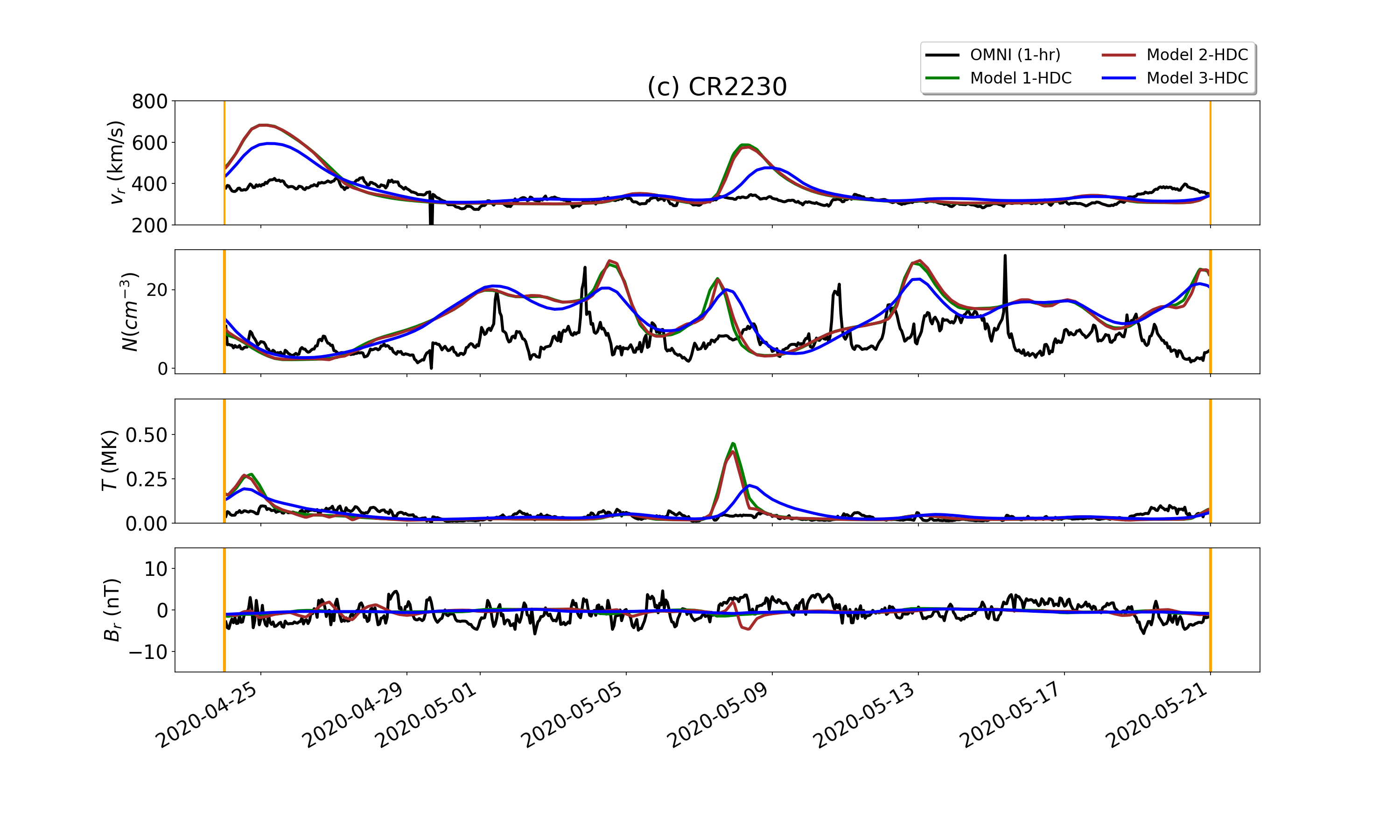}
     \caption{Comparison of Model 1 (green curves), Model 2 (brown curves), and Model 3 (blue curves) solutions using the HDC method for the radial velocity $V_{r}$ (km s$^{-1}$), radial magnetic field $B_{r}$ (nT), number density $N$ (cm$^{-3}$), and temperature $T$ (MK) with OMNI 1-hr in-situ measurements (black curves) for CR2215 (a), CR2220 (b), and CR2230 (c).}
    \label{fig:in-situ_comparisons_OMNI_HDC}
\end{figure*}

In Figure \ref{fig:in-situ_comparisons_OMNI_HDC}, we show the results for comparisons between the steady state SW solutions of Model 1 (green curves), Model 2 (brown curves), and Model 3 (blue curves) using the HDC method for the divergence-free condition with the OMNI 1-hr in-situ measurements (black curves) for three representative CRs: CR2215, CR2220, and CR2230. For instance, the results for all the variables for CR2215 are similar to the obtained with the Powell method for the same CR for the three models, as shown in the top panel of Figure \ref{fig:in-situ_comparisons_OMNI}. In the middle panel of Figure \ref{fig:in-situ_comparisons_OMNI_HDC}, we show the comparisons for CR2220; in this case, three models partially capture the observed rise of SW speed. In particular, three models underestimate the strength of the radial magnetic field, and for number density and temperature, both models overestimate the observed values. Interestingly, the low, medium, diffusive, and high diffusive models give comparable results, except in some sharp regions of density, similar to those obtained for the Powell eight-waves formulation. At the bottom of Figure \ref{fig:in-situ_comparisons_OMNI_HDC}, we display the results for CR2230, where it is discernible that three models behave similarly to the model using the Powell method. The models significantly overestimate the slow SW streams observed by OMNI and capture denser and hotter SW streams compared to in-situ measurements. Furthermore, three models underestimate the radial magnetic field at 1 AU, a typical issue of the numerical models. The latter results may suggest that this is to be expected for the solutions of MHD equations do not significantly modify the way of underestimating the observable magnetic field strength of the SW streams at 1 AU.

\subsection{Statistical analysis}
\label{Statistical_analysis}

We performed a statistical analysis to validate the quality of the three models and identify accuracy compared to in situ OMNI measurements. In particular, we compared measurements and model solutions by estimating the Normalized Root Mean Square Error (NRMSE) and the Pearson Correlation Coefficient (PCC). The NRMSE represents the normalized mean squared difference between measurements and models and facilitates comparing models with different scales \cite{REISS2022}, while the PCC measures linear correlation between two data sets. These two statistical measures are defined as follows:

\begin{eqnarray}
\textrm{NRMSE} &=& \frac{\sqrt{\frac{1}{N}\sum_{k=1}^{N}(f_{k}- O_{k})^{2}}}{(\textrm{max}(O_{k})-\textrm{min}(O_{k}))}, \label{NRMSE} \\
\textrm{PCC} &=& \frac{\sum_{k=1}^{N}(f_{k}-\bar{f_{k}})^{2}(O_{k}-\bar{O_{k}})^{2}}{\sum_{k=1}^{N}(f_{k}-\bar{f_{k}})^{2}\sum_{k=1}^{N}(O_{k}-\bar{O_{k}})^{2}}, \label{PCC} 
\end{eqnarray}

where ($f_{k}$, $O_{k}$) are the $k-th$ element of $N$ total model values and in situ observation pairs in the time series, i.e., it is a point-to-point comparison. The $\textrm{max}(O_{k})$ and the $\textrm{min}(O_{k})$ indicate the maximum and minimum values of the observed data, respectively. The $\bar{f_{k}}$ and $\bar{O_{k}}$ in equation \ref{PCC} represent the mean values.

Table \ref{Tab:statistical_results} contains the full statistical results represented by PCC/NRMSE of Models 1, 2, and 3 for the Powell method and HDC compared to OMNI's 1-hr data for the radial velocity $V_{r}$ in km s$^{-1}$, proton number density $N$ in cm$^{-3}$, proton temperature $T$ in MK and radial magnetic field $B_{r}$ in nT. 

\begin{table*}
\label{Tab:statistical_results}
\centering
\begin{tabular}{|c|c|c|c|c|c|c|c|}
\hline
    CR & Divergence-free method & Model & $V_{r}$ (km s$^{-1}$) & $B_{r}$ (nT) & $N_{p}$ (cm $^{-3}$) & $T_{p}$ (MK)      \\ \hline
 &    &  1  &  0.52/0.56  & 0.24/0.15  & 0.36/0.20  &  0.10/0.32     \\ 
CR2215      &  Powell eight-waves    &  2   & 0.51/0.55   & 0.24/0.15  & 0.37/0.20  &  0.10/0.33     \\ 
       &    &  3  &  0.48/0.47  & 0.29/0.14  & 0.42/0.20 &  0.14/0.26      \\ \hline
       &    &  1   &  0.51/0.56  &  0.23/0.15 &  0.36/0.21 &  0.09/0.34      \\ 
       & Cleaning    &  2  & 0.51/0.56  & 0.28/0.14 &  0.37/0.20 &  0.10/0.33     \\ 
       &    &  3   & 0.48/0.47  & 0.23/0.15 & 0.42/0.20 &  0.15/0.26     \\ \hline
 &    &  1  &  0.46/0.27  & 0.41/0.15  &  0.17/0.43 &  -0.05/0.22      \\ 
CR2220      &  Powell eight-waves    &  2   &  0.46/0.27  & 0.41/0.15  &  0.17/0.43 &   -0.04/0.22    \\ 
       &    &  3  &  0.45/0.25  & 0.46/0.15 & 0.12/0.42  & 0.02/0.20      \\ \hline
       &    &  1   & 0.45/0.28   & 0.38/0.15  & 0.16/0.43  &  -0.05/0.23     \\ 
       & Cleaning    &  2  & 0.46/0.27  & 0.24/0.15  & 0.17/0.43  &  -0.05/0.22     \\ 
       &    &  3   & 0.45/0.25  &  0.43/0.15 & 0.12/0.42  &   0.02/0.21    \\ \hline
 &    &  1  &  0.52/0.60  &  -0.09/0.21 & 0.24/0.32 &  0.23/0.70     \\ 
CR2230      &  Powell eight-waves    &  2   &  0.52/0.60  & -0.13/0.21  & 0.25/0.32  &  0.24/0.69     \\ 
       &    &  3  & 0.56/0.43  & -0.12/0.21  & 0.26/0.29  &  0.32/0.48      \\ \hline
       &    &  1   & 0.52/0.61   & -0.06/0.21 & 0.23/0.32  & 0.22/0.74      \\ 
       & Cleaning    &  2  &  0.52/0.60 &  -0.004/0.22 & 0.25/0.32  & 0.23/0.67      \\ 
       &    &  3   & 0.56/0.44  &  -0.08/0.21 &  0.26/0.29 & 0.32/0.48       \\ \hline
 &    &  1  & -0.34/0.34   & -0.08/0.11  & -0.06/0.25  & -0.14/0.09      \\ 
CR2250     &  Powell eight-waves    &  2   & -0.34/0.34  &  -0.07/0.11 &  -0.07/0.25 &  -0.13/0.09     \\ 
       &    &  3  &  -0.38/0.33 &  -0.06/0.11 & -0.07/0.25  & -0.14/0.09      \\ \hline
       &    &  1   &  -0.34/0.34  & -0.05/0.11  & -0.07/0.25 & -0.13/0.10       \\ 
       & Cleaning    &  2  & -0.35/0.34  &  -0.07/0.11 & -0.08/0.25  &  -0.14/0.10     \\ 
       &    &  3   & -0.38/0.33  & -0.03/0.11  &  -0.07/0.25 &  -0.15/0.10     \\ \hline
 &    &  1  &  0.50/0.33   & 0.55/0.17  &  0.29/0.24 & 0.13/0.17      \\ 
CR2260      &  Powell eight-waves    &  2   &  0.51/0.33  &  0.56/0.17 &  0.29/0.23 &  0.14/0.17     \\ 
       &    &  3  &  0.55/0.32  &  0.58/0.16 & 0.32/0.21  & 0.16/0.16      \\ \hline
       &  Cleaning  &  1   &  0.50/0.33  &  0.54/0.17 & 0.27/0.24 & 0.12/0.18       \\ 
       &    &  3   & 0.55/0.32  &  0.63/0.16 & 0.31/0.21 &  0.16/0.16     \\ \hline
\end{tabular}
\caption{Statistical Results of PCC/NRMSE for comparing Models 1, 2, and 3 using Powell eighth-waves and HDC with the OMNI 1-hr in-situ measurements.}
\end{table*}

\begin{figure*}
\centering
 \includegraphics[width=0.85\linewidth]
    {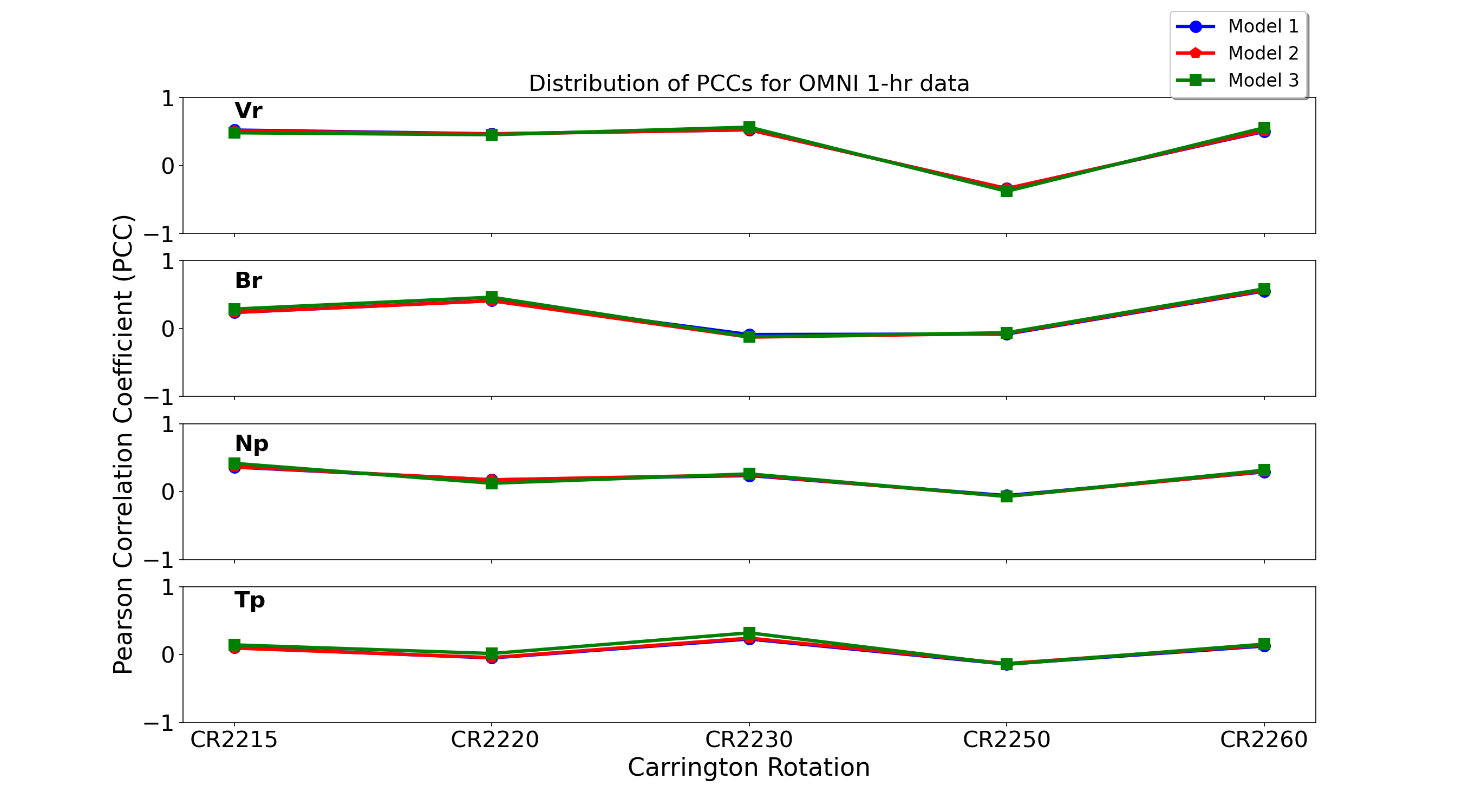}
    \includegraphics[width=0.85\linewidth]
    {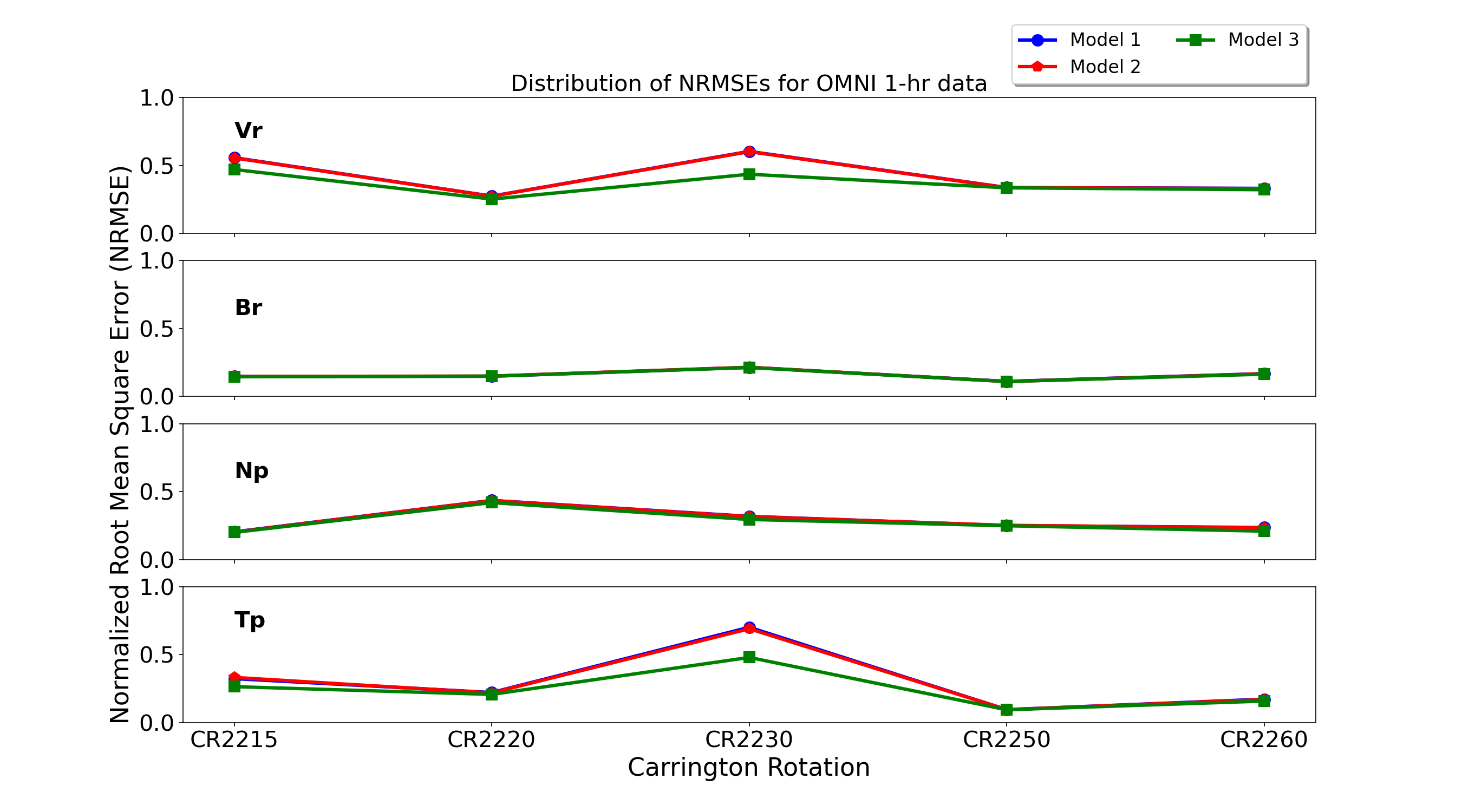}
     \caption{Statistical results of PCC (top) and NRME (bottom) of Models 1, 2 and 3 compared with OMNI 1-hr data.}
    \label{fig:statistical_analysis}
\end{figure*}

\begin{figure*}
\centering
 \includegraphics[width=0.85\linewidth]
    {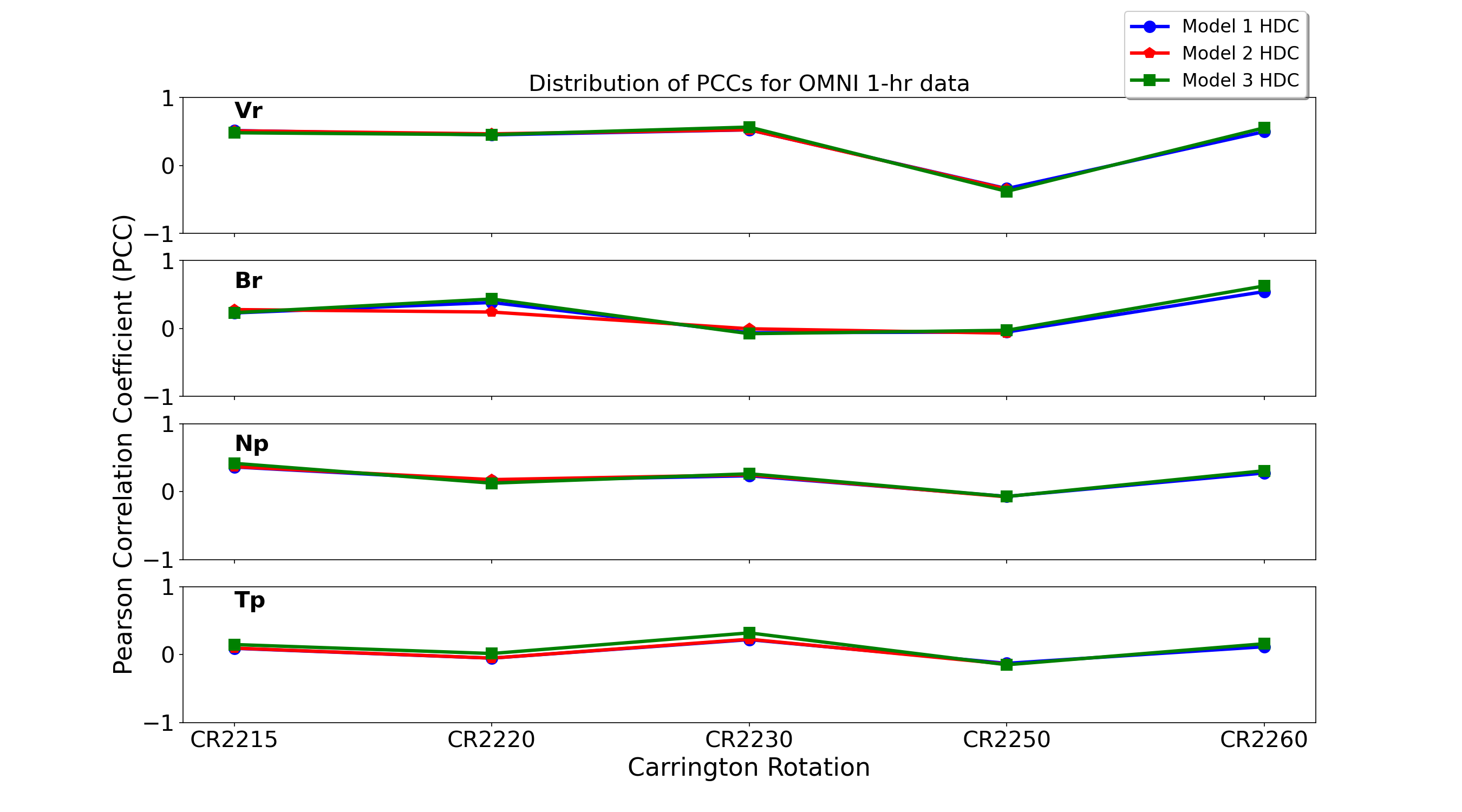}
    \includegraphics[width=0.85\linewidth]
    {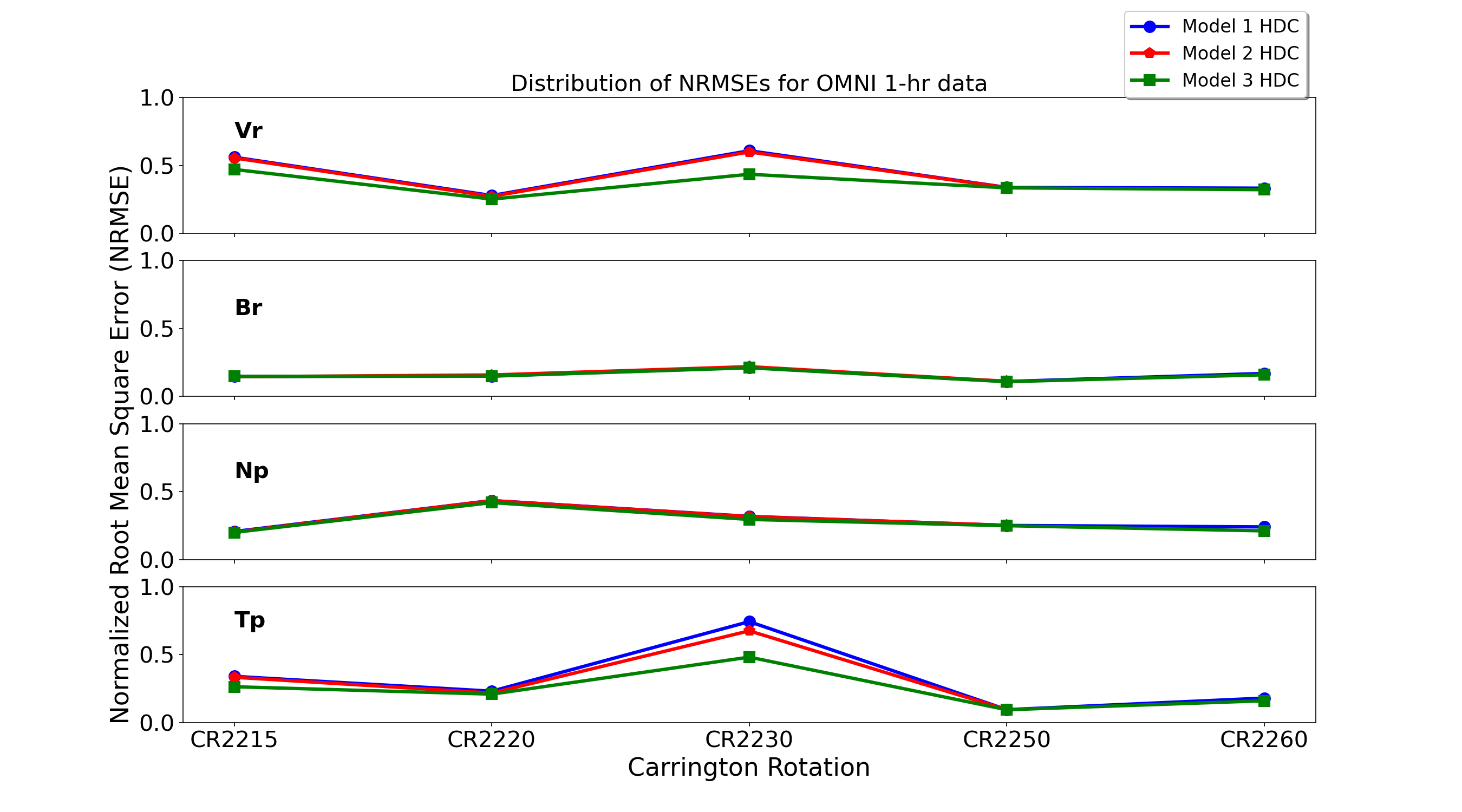}
     \caption{Statistical results of PCC (top) and NRME (bottom) of Models 1, 2, and 3 using the HDC method compared with OMNI 1-hr data.}
    \label{fig:statistical_analysis_HDC}
\end{figure*}

To identify more clearly the behavior of the statistical results of Table \ref{Tab:statistical_results}, in Figure \ref{fig:statistical_analysis}, we show plots of the results of PCCs and NRMSEs for radial velocity $V_{r}$, radial magnetic field $B_{r}$, proton number density $N_{p}$ and proton temperature $T_{p}$ using the Powell method in the comparisons between models and OMNI measurements for all the CRs considered in this paper.  Each point of Figures \ref{fig:statistical_analysis} and \ref{fig:statistical_analysis_HDC} represents the value of the PCC and NRMSE calculated by a point-to-point comparison using the equations (\ref{NRMSE}) and (\ref{PCC}) as explained above for each variable for each CR.

At the top of Figure \ref{fig:statistical_analysis}, we display the distribution of PCCs, where it is discernible that the best correlations are shown for the radial velocity and radial magnetic field, especially, Model 3 gets the highest value of the whole analysis for the radial velocity, that is, PCC = 0.56 for CR2230. The three models get negative PCCs for CR2250, i.e., simulation results and OMNI observations tend to be anti-correlated, but for this CR, we should consider that steady-state SW solutions could be poor due to transient events such as CMEs. A remarkable result is that the three models have very similar values of PCC for all the variables and all the CRs, i.e., the plots practically overlap. The latter implies that using robust numerical methods to achieve acceptable PCC values with in-situ measurement at 1 AU is not strictly necessary. At the bottom of Figure \ref{fig:statistical_analysis}, we display the distribution of NRMSEs, which look similar for the three models and CRs. Surprisingly, for most CRs, Model 3 performs better regarding NRMSE than Models 1 and 2 for all the SW parameters. 

In Figure \ref{fig:statistical_analysis_HDC}, we show plots of the results of PCCs and NRMSEs for radial velocity $V_{r}$, radial magnetic field $B_{r}$, proton number density $N_{p}$ and proton temperature $T_{p}$ using the HDC method obtained in the comparisons between models and OMNI-1 hr measurements for all the CRs considered in this paper. In the two panels of Figure \ref{fig:statistical_analysis_HDC}, we note that the distribution of PCCs and NRMSEs are comparable to those shown in Figure \ref{fig:statistical_analysis} for the case of the Powell method. However, some values slightly vary; for example, the highest PCC was 0.63, corresponding to Model 3 in the case of the radial magnetic field for CR2260. In the case of NRMSE, the lowest values were achieved for CR2215 and CR2220 for the radial magnetic field. Overall, the values of PCC and NRMSE for both divergence-free methods (Powell and HDC) are comparable.

\section{Conclusions}
\label{conclusions}

In this paper, we have used sunRunner3D to explore the effects of numerical methods on global MHD simulations of steady-state SW solutions in the inner heliosphere. According to the results for the three models for both divergence-free methods (Powell and HDC) for the selected CRs, we find that sunRunner3D is capable of capturing the typical properties of the SW streams in the inner heliosphere, including the CIRs and the Parker Spiral structures. Specifically, Model 1 for both divergence-free methods shows more developed CIRs. While Model 2 shows comparable solutions to Model 1, Model 3, which has a highly diffusive combination of algorithms, obtains more diffusive structures than Models 1 and 2. Despite the diffusion, the three models for the Powell and HCD methods capture the typical SW features in the inner heliosphere. 

Regarding the comparison of the models with OMNI-1hr in-situ measurements using the Powell and HDC methods, the statistical analysis shows that the values of the three models are acceptable regarding the PCC and NRMSE for the selected CRs. The three models get high PCC values for radial velocity and radial magnetic field for most CRs except for CR2250. On the other hand, the three models get the lowest PCC values for number density and temperature, which could be related to the use of an ideal equation of state. In the case of NRMSE, again, the three models obtained the lowest values, i.e., the best solutions, for radial magnetic field for most of the CRs, while the highest NRMSE values, i.e., the worst solutions are for radial velocity and temperature for CR2215 and CR2230. On the other hand, the results of the comparisons of the three models with OMNI-1 hr observations for the case of the HDC method are overall comparable with those obtained with the Powell method, which could indicate that the use of any divergence-free condition method leads to similar results regarding the steady-state solutions of the SW in the inner heliosphere. A relevant result in the statistical analysis is that Model 3, i.e., the highly diffusive model, behaves better than Models 1 and 2 for most of the variables in all the CRs. The latter result raises the question of using a low diffusive, which could be more computationally demanding, or a high diffusive, which is less time-consuming. However, we should also consider other parameters that we do not consider in this paper, such as spatial resolution, which could increase the computational time and necessitate the use of supercomputers. Also, this paper does not study other divergence-free condition methods, such as constrained transport, which has already been used for global MHD SW simulations in the heliosphere \cite{Pomoell&Poedts_2018}. However, we need to define staggered variables from the boundary conditions of CORHEL (which are cell-centered) to implement it properly into PLUTO. Nevertheless, the spatial resolution and the divergence-free methods used here have produced consistent results comparable with those of the state of the art regarding simulations of SW streams in the inner heliosphere. 
Furthermore, as the primary purpose of this paper is to investigate the effects of numerical algorithms, we also decided to study different methods to numerically maintain the $\nabla\cdot{\bf B}$ close to zero. These two methods are available in the MHD module of the PLUTO code and are naturally consistent with the finite volume method adopted to solve the ideal MHD equations. Also, the analysis of different divergence-free methods has rarely been studied in the context of global MHD simulations of SW in the heliosphere. Therefore, including this analysis in our paper represents a novel study that could be helpful for the space physics community modeling when deciding the most appropriate algorithm combination. Besides, it would be interesting to test the constraint transport method because it has been used by other models in global MHD simulations of the SW as in \cite{Pomoell&Poedts_2018}. This method maintains $\nabla\cdot{\bf B}$ around the machine round-off error. Unfortunately, it is not straightforward to implement in spherical coordinates due to technical issues such as the definition of a staggered mesh so that we might consider it for a future study. In summary, since the SW magnetic field is predominantly radial, the divergence-free methods (Powell and HCD) employed in this paper control $\nabla\cdot{\bf B}$ to a truncation error, but this does not affect the global structure of the SW streams in the heliosphere and their comparisons with in situ measurements of the OMNI data.

Finally, according to the steady-state solutions of the SW, the comparisons between the models and OMNI-1hr in-situ measurements, and the statistical analysis, we can conclude that the numerical methods related to the solutions of MHD equations using finite volume do not significantly affect the global MHD simulations of steady-state SW in the inner heliosphere. Notably, the diffusion level only affects the structures of the SW streams and slightly modifies the morphology. Nevertheless, it does not affect the accuracy of the physical parameters associated with the SW since the comparisons with in-situ measurements show similar results for the three models with the two divergence-free methods despite the diffusion in sharp regions of number density or high-speed SW streams. Therefore, this paper's results could encourage using less robust numerical methods that imply fewer computational resources but without losing accuracy in global MHD simulations of steady-state SW streams in the inner heliosphere. Also, our results suggest that the simulations should not necessarily run on supercomputers. For example, Model 1 for CR2215 with the Powell method lasts about 3 hours using 32 cores on a workstation, while Model 3 lasts around 2 hours in the same workstation. Then, we could contemplate the use of a decent workstation (at least with 40 cores) to obtain consistent results of the global behavior of the SW streams and, going further, we could make forecasting of the SW conditions near the Earth environment ($\sim$ 1 AU) quickly, which might be meaningful for Space Weather applications.

\section*{Acknowledgments}
We appreciate the referee's careful review and constructive suggestions that helped to increase the draft's clarity. J.J.G.-A. acknowledges the "Consejo Nacional de Humanidades Ciencias y Tecnolog\'ias (CONAHCYT)" 319216 project "Modalidad: Paradigmas y Controversias de la Ciencia 2022," for the financial support of the research carried out in this work. In particular, L.A.L.-A acknowledges the scholarship received as part of a bachelor student associated with the 319216 project. We carry out the numerical simulations in the cluster purchased with resources of the 319216 project. We generate Figure \ref{fig:solar_wind_relaxation_3D_CR2215_Model_1} using VisIt (\href{https://visit-dav.github.io/visit-website/index.html}{https://visit-dav.github.io/visit-website/index.html}). All the authors are grateful to the developers of the PLUTO software, which provides a general and sophisticated interface for the numerical solution of mixed hyperbolic/parabolic systems of partial differential Equations (conservation laws) targeting high Mach number flows in astrophysical fluid dynamics.

\section{References}

\noindent



\medline
%
\nocite{*}
\bibliographystyle{rmf-style}
\bibliography{Investigating_effects_numerical_algorithms_solar_wind_MHD_simulations_RMF_2024_accepted_version}
%
%
%

%
%

\end{document}